\documentclass[aps,pre,twocolumn,showpacs,floatfix,superscriptaddress]{revtex4-2}
\usepackage{subfigure,amsmath, amsthm, amssymb}
\usepackage{graphicx}
\usepackage{graphics}
\usepackage[export]{adjustbox}
\usepackage[usenames]{color}
\usepackage{float}
\usepackage{physics}
\usepackage{tabularx,ragged2e}
\usepackage[geometry]{ifsym}
\usepackage{wrapfig,lipsum,booktabs}
\newcommand{\be}{\begin{equation}}
\newcommand{\ee}{\end{equation}}
\newcommand{\bea}{\begin{eqnarray}}
\newcommand{\eea}{\end{eqnarray}}

\begin{document}
\title{Kinetic model description of dissipation and recovery in collagen fibrils under cyclic loading}
\author{Amir Suhail}
\email{amirs@imsc.res.in}
\affiliation{The Institute of Mathematical Sciences, CIT Campus, Taramani, 
Chennai 600113, India}
\affiliation{Homi Bhabha National Institute, Training School Complex, Anushakti Nagar, Mumbai 400094, India}
\author{Anuradha Banerjee}
\email{anuban@iitm.ac.in}
\affiliation{Department of Applied Mechanics, IIT-Madras, Chennai-600036, India}
\author{R. Rajesh}
\email{rrajesh@imsc.res.in}
\affiliation{The Institute of Mathematical Sciences, CIT Campus, Taramani, 
Chennai 600113, India}
\affiliation{Homi Bhabha National Institute, Training School Complex, Anushakti Nagar, Mumbai 400094, India}
  

\begin{abstract}
Collagen fibrils, when subjected to cyclic loading, are known to exhibit hysteretic behaviour with energy dissipation that is partially recovered on relaxation. In this paper, we develop a kinetic model for a collagen fibril incorporating presence of hidden loops and stochastic fragmentation as well as reformation of sacrificial bonds. We show that the model reproduces well the characteristic features of reported experimental data on cyclic response of collagen fibrils, such as moving hysteresis loops, time evolution of residual strains and energy dissipation, recovery on relaxation, etc. We show that the approach to the steady state is controlled by a characteristic cycle number for both residual strain as well as energy dissipation, and is in good agreement with reported existing experimental data.
\end{abstract}

\maketitle

\section{\label{introduction} Introduction}

Collagen, the most abundant protein in humans, is found in several hard and soft tissues, such as, bones, tendons, ligaments, cartilage etc. Collagenous tissues provide not only mechanical support and strength but also flexibility and mobility~\cite{Kalder2007, Fratzl2008,Shoulders2009}. Collagen has a hierarchical structure with tropocollagen as the fundamental protein molecule. The diversity observed in the mechanical behavior of collagen based tissues is a direct consequence of the differences in their hierarchical structures~\cite{OTTANI2002,FRATZL20071263}. Characterizing structure-property  of collagen at different length scales and developing predictive models have significance not only in understanding the mechanistic basis of wide spectrum of properties seen in different tissues but also for important clinical objectives, such as risk assessment of tissue failure, treatment optimization, etc~\cite{parenteau2010collagen,lee2001biomedical}.

Collagen molecules, typically of length $\approx$ 300 nm, self assemble in a staggered manner to form long collagen fibrils of diameters ranging between 10s to 100s of nm~\cite{Fratzl2008}. The staggered arrangement of these molecules in the longitudinal direction results in a gap and overlap region in the fibril along its length which gives rise to the characteristic D-period of the fibril~\cite{PETRUSKA1964,Orgel2006}. The fibril structure is further stabilized by intermolecular enzymatic covalent cross-linking that form at the non-helical ends (telopeptides)~\cite{Light1980,KNOTT1998,Reiser1992}. 

At the smallest length scale, the mechanical response of collagen molecules has been determined using atomic force microscopy (AFM) and optical tweezer  experiments~\cite{Thompson2001,SUN2002,SUN2004,BOZEC2005}. The molecular basis of toughness of collagenous tissues was established by identifying the basic mechanisms of energy dissipation during pulling of collagen molecules using AFM~\cite{Thompson2001}. The force-extension response of collagen from bovine Achilles tendon was shown to be saw-toothed, such that the force had multiple drops with increasing extension. These drops were attributed to the rupture of intermolecular sacrificial bonds that release hidden lengths, thereby ensuring the integrity of the backbone chain and at the same time dissipating large amounts of energy. Further, a delay of 100s before the next cycle was shown to result in almost $50\%$ recovery in the capacity of energy dissipation, suggesting possible reformation of the sacrificial bonds during the waiting interval~\cite{Thompson2001}. Similar saw-toothed response was also seen in molecular scale experiments of biological polymeric adhesive found in nacre~\cite{Smith1999}, in unfolding of titin~\cite{Reif1997} etc. The structure of intermolecular covalent cross-linking between adjacent molecules, using X-ray diffraction, was shown to have a turn at the C-terminal telopeptides that causes the molecule to fold back on itself~\cite{ORGEL2001}.

At the level of fibrils, the extent and type of covalent cross-linking between the tropocollagen molecules has been shown to strongly affect the constitutive response~\cite{SVENSSON20132476,SHEN2010}. The collagen fibrils from human patella tendon were found to exhibit a characteristic three phase stress-strain behavior. An initial rise in modulus followed by a plateau and in the final phase further increase in stresses and modulus, hypothesized to be a consequence of maturity of cross-links,  before final failure~\cite{SVENSSON20132476}. In contrast, collagen from rat tail tendon, a non-load bearing tissue, displayed only two phases as plateau in the stress-strain led to failure.

Simulations at multiple length scales have provided interesting insights into various aspects of deformation and failure of collagen ranging from atomistic length scales, focusing on individual tropocollagen molecules, to continuum length scales for collagen fibrils~\cite{LORENZO2005,Buehler2006a,Buehler12006b}. Atomistic-scale investigations showed three stages of tropocollagen deformation: molecular unwinding, breaking of hydrogen bonds and backbone stretching and the overall response was rate dependent~\cite{GAUTIERI2009}. Mesoscopic molecular model,  derived from atomistic studies of tropocollagen, of ultra-long tropocollagen molecule  showed the transition from entropic elasticity at small deformations to energetic elasticity at large deformations~\cite{BuehlerWong2007}. Atomistic simulations of Uzel et. al~\cite{UZEL2011}, incorporating the folded structure of the cross link as seen using X-ray diffraction~\cite{ORGEL2001}, were shown to reproduce the dissipative response of collagen molecules better than the earlier studies. The observed unfolding of non-helical regions, as well as the stretching and eventual breakage of enzymatic cross-links can be equated to the breaking of sacrificial bonds and the release of hidden length. 

At the level of fibrils, using idealized two-dimensional representation of collagen fibril, large deformations without catastrophic failure were shown to be possible due to molecular stretching as well as other competing mechanisms such as intermolecular sliding and breaking of cross-links between collagen molecules~\cite{BUEHLER2007}. More realistic aspects of structure of collagen were incorporated in a three dimensional model of fibril~\cite{DEPALLE20151}, with enzymatic cross-links included as in their physiological locations. By differentiating between mature and immature cross-link properties, the model could reproduce the three-phase stress-strain response as in experiments~\cite{SVENSSON20132476}. A similar molecular dynamics (MD) simulation of a three dimensional model showed the effect of degradation in properties of cross-links  at the fibril surface or within the volume on the overall fibril response~\cite{MALASPINA2017549}. With small degradation, a drastic change in mechanical properties was observed, demonstrating the relevance of molecular organization in collagen fibrils.

Collagen is subjected to cyclic loads during exercise and routine body movements. While the response of collagen to monotonically increasing loads is comprehensively investigated, the response to cyclic loads, resulting dissipation and recovery are comparatively much less studied~\cite{SHEN2008cyclicload,SVENSSON2010,LIU2018cyclicload}. Shen et. al.~\cite{SHEN2008cyclicload} performed fatigue test on isolated collagen fibrils and reported four different stress-strain response: linear to failure, perfectly plastic, perfectly plastic-strain hardening, and nonlinear strain softening. All fibrils exhibited significant hysteresis and a residual strain (strain at zero force). A recovery in residual strain was also observed, which was dependent on the amount of time spent at zero force.

In a recent study, Liu et. al.~\cite{LIU2018cyclicload} conducted displacement controlled cyclic loading experiments on single collagen fibrils obtained from calf skin. Collagen fibrils were loaded for 20 cycles up to a predetermined stretch ratio, $\lambda^{max}$, and then unloaded till zero force. The fibrils were allowed to relax for 1 hour after the first 10 cycles. The stress-stretch response of fibrils showed moving hysteresis loops and associated residual strains. With increasing number of loading cycles, the dissipation during hysteresis decreases while  the residual strain increases and both finally saturate to their respective steady state values. Collagen fibrils also showed recovery in residual strain and as well as in capacity to dissipate energy when allowed to relax at zero force. It was conjectured that these features could be due to the existence of reformable sacrificial bonds within the fibrils. Finally, the fibrils which were cyclically loaded showed an increase in strength and toughness, compared to monotonically loaded fibrils. The mechanism underlying these enhancements was speculated to be due to some permanent molecular rearrangements. With respect to modeling cyclic response of collagen, there are recent advances in understanding of the energy dissipation and wave propagation properties of collagen at molecular and microfibril level due to transient loading using fully atomistic models [28–30]. However, to the best of our knowledge, existing models have not explained key experimental features from cyclic loading of a single collagen fibril.

Dynamic sacrificial bonds within polymers have been successfully incorporated in simplified models called kinetic models. The saw-toothed stress-strain response of collagen molecules has been  simulated using deterministic kinetic models of a worm-like chain with additional sacrificial bonds whose breakage results in the release of a hidden length, resulting in a drop in force~\cite{elbanna2013dynamics,Lieou2013pre}. Historically,  two state kinetic models have been used to describe the force-extension response of single protein pulling experiments~\cite{Reif1998,SU2009}. In this paper, within the framework of kinetic models, we develop a minimal stochastic kinetic model for collagen fibrils that incorporates dynamic reformable sacrificial bonds with hidden lengths. We show that the proposed model is able to reproduce the main qualitative features of the cyclic loading experiment~\cite{LIU2018cyclicload}, suggesting that the essential physics is captured by the  kinetic model. By choosing realistic model parameters, we reproduce key quantitative features of the experimental data.

The remainder of paper is organized as follows. In Sec.~\ref{Method}, we  describe the stochastic kinetic model and its implementation for collagen fibril. In Sec.~\ref{Results}, we determine the stress-stretch response of the fibril, the recovery of the fibril on relaxation, and the behavior of the dissipation as well as residual strain with cycles. We show that our model reproduces main features of cyclic loading experiment. Section~\ref{conclusion} contains a summary and discussion of the results.

\section{\label{Method} Model and Methods}

\subsection{Model}

{\bf Kinetic model formulation:} We first describe the basis of kinetic models and how they incorporate the dynamic formation and breaking of sacrificial bonds. We then give the details of the specific kinetic model that we develop for simulating the cyclic response of a fibril.

\begin{figure}
	\centering
\includegraphics[width=\columnwidth, frame]{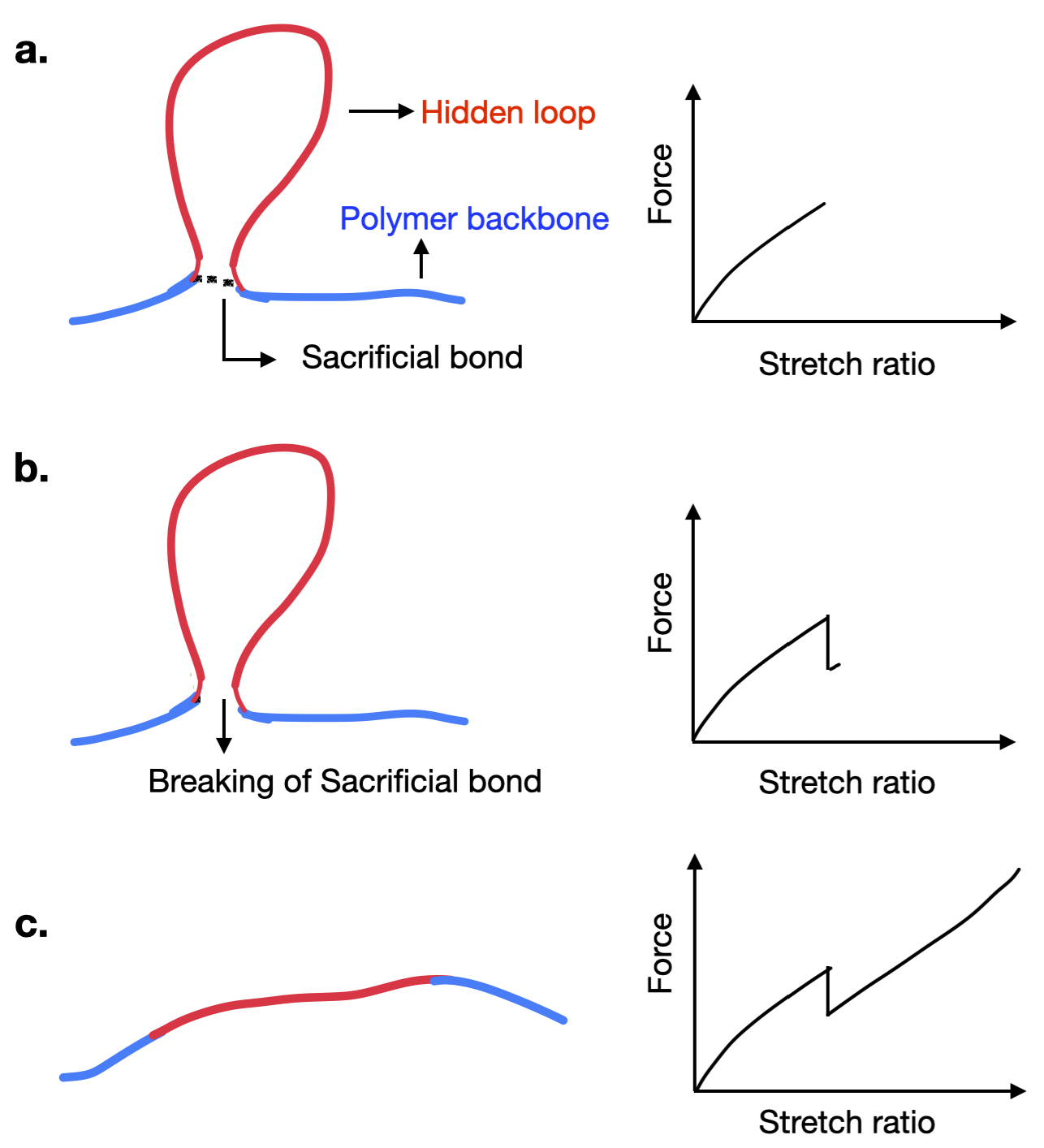}
	\caption{(a) Schematic of a polymer with a single sacrificial bond (dotted line), corresponding hidden loop (shown in red) and the corresponding force-stretch ratio response. (b) As the sacrificial bond breaks, the force drops due to release of the hidden length. (c) Force rises again as the polymer is extended further.}
	\label{fig:1}
\end{figure}

Consider a linear polymer whose contour length, in the absence of sacrificial bonds, is $L_c$. Let bond length be $b$ such that number of monomers are $N=L_c/b$. Each sacrificial bond creates a hidden loop that prevents a part of polymer backbone from taking any load, as shown schematically in Fig.~\ref{fig:1}. When hidden loops are present, the available length $L_a$, of the polymer backbone is less than $L_c$ and is given by 
\begin{equation} \label{eq:1}
L_a=L_c-\sum_{i} \ell_i,
\end{equation}
where $\ell_i$ is the length of the $i$th hidden loop. The length of the hidden loops are chosen from a distribution $P(\ell)$. 

We denote the  stress-stretch relation of the polymer  by $\sigma(\lambda)$, where $\lambda$ is the stretch (note that $\lambda=1+\epsilon$, where $\epsilon$ is the strain). We assume that $\sigma(\lambda)$ increases monotonically with $\lambda$. Sacrificial bonds are created and broken with rates $k_b$ and $k_f$ which are in general dependent on the force acting on the polymer. For a given macroscopic extension, when a sacrificial bond is created, $L_a$ decreases, thus increasing the strain, and hence the force. Similarly,  when a sacrificial bond breaks, $L_a$ increases,  thus decreasing strain, and hence there is a drop in force. The rates of formation, $k_b$, and fragmentation, $k_f$, of sacrificial bonds have been earlier modeled~\cite{Lieou2013pre}, according to Bell's theory~\cite{Bell}, as
\bea
 \label{eq:2}
 k_f&=&\alpha_0 \exp \left (\frac{F\Delta x_f}{k_BT} \right),\\
 \label{eq:kb}
k_b&=&\beta_0 \exp \left (\frac{-F\Delta x_b}{k_BT} \right),
\eea
where $\alpha_0$ and $\beta_0$ are rates of fragmentation and formation of sacrificial bonds at zero force, $\Delta x_f$ and $\Delta x_b$ are distances to transition state, $F$ is the force felt by the sacrificial bond, $k_B$ is the Boltzmann's constant and $T$ is the temperature.
\begin{figure}
	\graphicspath{{images/}}
	\centering
	\includegraphics[width=\columnwidth]{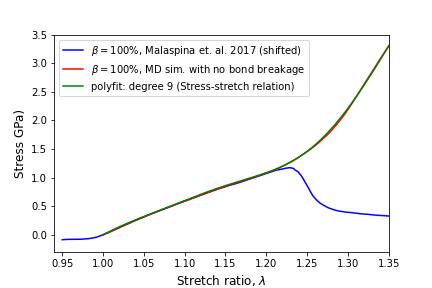}
	\caption{Stress-stretch relation $\sigma(\lambda)$ obtained from MD simulations of a fibril in which breakage of  bonds (including backbone and other enzymatic cross-links) is disallowed. The data is fitted to a polynomial of degree nine. The $x$-axis has been shifted to ignore the knee region.}
	\label{fig:2}
\end{figure} 

{\bf Determination of stress-stretch relation:} We now describe the implementation of the kinetic model for a collagen fibril. A fibril consists of a collection of collagen molecules that are linked to each other through enzymatic crosslinks. Within the kinetic model framework, we treat the collagen fibril as a coarse grained linear polymer.  The crosslinks are treated as dynamic sacrificial bonds that can be created or broken with rates described in Eqs.~(\ref{eq:2}) and (\ref{eq:kb}). To first establish the stress-stretch response of collagen fibril without any creating or fragmentation dynamics, we use an existing coarse grained three dimensional MD model~\cite{MALASPINA2017549}, but here we disallow any fragmentation of crosslinks. In the MD model, each collagen molecule is represented by a linear bead-spring model of 215 beads. To create a microfibril, five collagen molecules are arranged in a staggered manner along the longitudinal direction and as a pentagon in the transverse direction. Repeated arrangement of multiple microfibrils forms a fibril. The different molecules are connected to each other through cross-links within each microfibril. A detailed description of the model, the values of the parameters used, and details of simulation are provided in Appendix~\ref{appendix:a}.

The stress-stretch  relation $\sigma(\lambda)$, where $\lambda=x/L_a$ and $x$ is the end-to-end distance of the polymer, obtained from MD simulations  is shown in Fig.~\ref{fig:2}, where for bench-marking, we have compared the data with the results of Ref.~\cite{MALASPINA2017549}, where crosslinks break beyond a threshold strain. For convenience of use in the kinetic model, we fit a ninth order polynomial 
\begin{equation}\label{eq:3}
	\sigma(\lambda)=\sum_{n=1}^{9} a_n (\lambda-1)^n 
\end{equation}
to the data.
\begin{figure}
	\graphicspath{{images/}}
	\centering
		\includegraphics[width=0.5\textwidth]{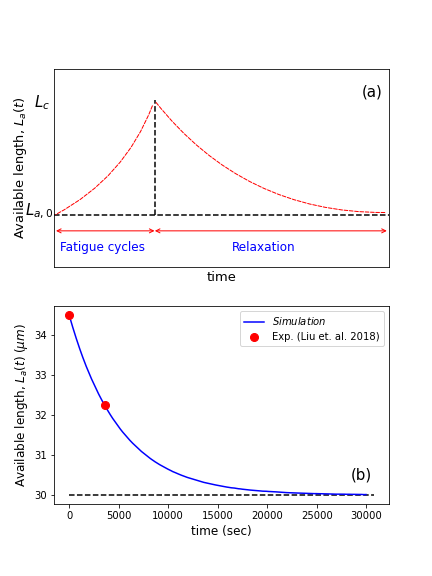} 
	\caption{(a) Schematic of evolution of available length during a series of cycles followed by relaxation at zero force. (b) Relaxation dynamics of fibril of length $L_c$(with no sacrificial bonds). At long time fibril equilibrates to initial experimental length $L_{a,0}$. Relaxation curve averaged over 1000 runs.}
	\label{fig:3}
\end{figure}

{\bf Determination of parameters and rates:} We now describe how to determine model parameters: $L_c$, $P(\ell)$, $\alpha_0$ and $\beta_0$. At zero force,  sacrificial bonds form and break spontaneously with rates $\beta_0 L_a/b$ and $\alpha_0 N_b$ respectively, where $N_b$ is the number of sacrificial bonds present at any instant. At steady state, rate of fragmentation and formation of bonds should be equal, implying
\begin{equation}\label{eq:6}
	\alpha_0 \langle N_b \rangle  = \frac{\beta_0 \langle L_{a, 0} \rangle}{b}, ~~F=0,
\end{equation}
where the zero in the subscript of  $L_{a,0}$ denotes the reference time after steady state is reached, taken to be $t=0$.
Also, $\langle L_{a,0}\rangle =L_c-\langle N_b \rangle \langle \ell \rangle$ where $ \langle \ell \rangle$ and $ \langle N_b \rangle$ are the average loop size and the average number of loops respectively. Substituting for $\langle L_{a,0} \rangle$ in Eq.~(\ref{eq:6}), we obtain
\bea
\langle N_b \rangle &= &\frac{L_c \beta_0}{\alpha_0 b + \beta_0 \langle \ell \rangle}, \label{eq:8}\\
\frac{L_c}{L_{a, 0}} &=&1+\frac{\beta_0}{\alpha_0} \frac{\langle \ell\rangle}{b}.
\label{eq:9}
\eea

We estimate $L_c/\langle L_{a, 0}\rangle$ from the experimental data \cite{LIU2018cyclicload}. To do so, we assume that after 20 cycles, most of the sacrificial bonds are broken. Equating the ratio $L_c/ \langle L_{a,0} \rangle$ to the experimental residual extension of $\approx 1.15$ after $20$ cycles, as shown in Fig. 5(c) of Ref.~\cite{LIU2018cyclicload}, we obtain $L_c/\langle L_{a, 0}\rangle=1.15$. The initial length of the fibril is known to be $\langle L_{a, 0}\rangle=30\; \mu m$, thus fixing $L_c$. The inter-monomer distance $b$ is chosen to be $b=1.4$ nm, equal to the  inter-bead distance in the MD model~\cite{MALASPINA2017549}. To choose the distribution of the hidden loop sizes, we proceed as follows. In kinetic models for collagen, the loop sizes were chosen proportional to the contour length of the polymer. However, in fibrils, the sacrificial bonds (cross-links) are formed between neighboring monomers of different molecules and involve utmost a few monomers. We therefore expect the loop size to be the order of four monomer lengths. Hence, we make the choice of $P(\ell)$ to be a uniform distribution $U[2b,6b]$. We will argue that this choice is consistent with the MD-model for fibrils, as well as show that the results are not sensitive to the choice as long as the perturbations to $P(\ell)$ are not significant. With this choice of $P(\ell)$, we obtain $\langle \ell \rangle =4 b$.

On substituting these values  of $L_c/\langle L_{a,0} \rangle$, $b$ and $\langle \ell \rangle$ in Eq.~(\ref{eq:9}), we obtain $\beta_0/\alpha_0=0.0375$. Then from Eq.~(\ref{eq:8}), we obtain $\langle N_b \rangle/ (L_c/b) \approx 0.0326$. We now argue that this number that follows from the experimental residual strain has the correct order of magnitude. The kinetic model represents a fibril with diameter of a single microfibril, such that 215 monomers in the kinetic model represents $215\times 5$ monomers of the microfibril. A molecule in the microfibril has two crosslinks. This corresponds to 10 sacrificial bonds per 215 monomers in the kinetic model or equivalently we expect $\langle N_b \rangle/ (L_c/b) \approx 0.047$. Among these, some will be broken at zero force, and the calculated result  $\langle N_b \rangle/ (L_c/b) \approx 0.0326$ makes sense. 

Knowing the ratio $\beta_0/\alpha_0=0.0375$, we would like to now fix the values of $\alpha_0$ and $\beta_0$. For this, we use the fact that as part of the cyclic loading experiment~\cite{LIU2018cyclicload},  recovery of residual strain is also studied.  In the experiment, the fibril is cyclically loaded for 10 cycles followed by relaxation at zero force for 60 minutes, as shown schematically in Fig.~\ref{fig:3}(a). We will choose an $\alpha_0$ for which the relaxation time matches with the experimental data. For doing so, we take a  polymer of length $L_c$  with no sacrificial bonds which roughly mimics the state after 10 cycles. We then equilibrate the system at zero force. After equilibration, the available length is $L_{a_0}$, as shown in Fig \ref{fig:3}(b). The relaxation dynamics from our model matches well with the experiment (experimental data shown as solid circles)  for $\alpha_0= 1.6853 \times 10 ^{-4} s$, as shown in Fig \ref{fig:3}(b). 

Finally, we describe how we fix the parameters $\Delta x_f$ and $\Delta x_b$, as defined in Eqs.~(\ref{eq:2}) and (\ref{eq:kb}). The force $F$ in these equations is the force felt by the sacrificial bonds. Since the sacrificial bonds or crosslinks are between different collagen chains and transverse to the direction of loading, we have no direct way of measuring $F$. Instead, we approximate it by the force in a chain. In the MD simulations, the force in a chain is $\sigma A_0/185$, where $A_0$ is the cross-sectional area of the fibril, and $185$ is the number of chains. We then treat $\Delta x_f$ as a parameter. Note that $\Delta x_f$ controls when the stretch ratio at which fragmentation of sacrificial bonds is enhanced. We perform a parametric study of the dependence of the stress-stretch response for uniaxial loading on $\Delta x_f$. We choose that value of $\Delta x_f$ for which the strain at which deviation from the initial linear behavior coincides with that in the experiment. Using this procedure, we converge on  $\Delta x_f$ to be $.01$ nm. We notice that the formation rate is low and during the pulling experiment, there are very few reformations of sacrificial bonds. We therefore choose $\Delta x_b$ to be zero, and check that even if a non-zero value is chosen, the results do not change.

The values of the different parameters are summarized in Table~\ref{table:t1}.
\begin{table}
	\caption{\label{table:t1} The parameters for the kinetic model for collagen fibril. }
	\begin{ruledtabular}
		\begin{tabular}{l p{4.0cm}l}
			Parameter&Description & Value \\
			\hline
			$L_{a, 0}$& available length at zero force& $30\; \mu m$ \\
			$L_c$ &contour length  & $ 1.15\; L_{a,0}$ \\
			$b$ & bond length  &$1.4\; nm$ \\
			$P(\ell)$ &loop size distribution & $U[2b,6b]$ \\
			$\langle \ell \rangle$ & mean loop size & 4b \\
			$\beta_0$& formation rate of sacrificial bonds at zero force& $6.32\times 10^{-6} s^{-1}$  \\
			$\alpha_0$& fragmentation rate of sacrificial bonds at zero force  & $1.69 \times 10 ^{-4} s^{-1}$ \\
			$\Delta x_f$& distance to transition state & $.01\;nm$ \\
			$\Delta x_b$& distance to transition state & 0  \\
			$v$& pulling velocity& $125\; nm/s$\\
			$T$ &temperature &  $298\;K$ 
		\end{tabular}
	\end{ruledtabular}
\end{table}

\subsection{Simulation Protocol \label{sec:protocol}}

The system evolves in time through constant time steps $dt$. In this time interval,  the probabilities of fragmentation ($p_f$) and formation ($p_b$) of sacrificial bonds are given by $p_f=k_f N_b(t) dt$ and $p_b=k_b N_f(t)dt$
where, $N_f(t)=L_a(t)/b$ is the number of free sites and $N_b(t)$ is the number of sacrificial bonds. The time step $dt$ is chosen such that the probabilities are  much smaller than $1$ at all times. Whenever a sacrificial bond forms, a hidden loop of length $\ell$ is assigned from distribution $P(\ell)$. When a sacrificial bond breaks, a hidden length of a randomly chosen loop is released. The available length gets updated as $L_a \pm \ell$ depending on breaking/formation event of sacrificial bonds.  The rates are also updated depending on the current force and current $L_a$.
\begin{figure}
	\graphicspath{{images/}}
	\centering
	\includegraphics[width=1.0 \linewidth]{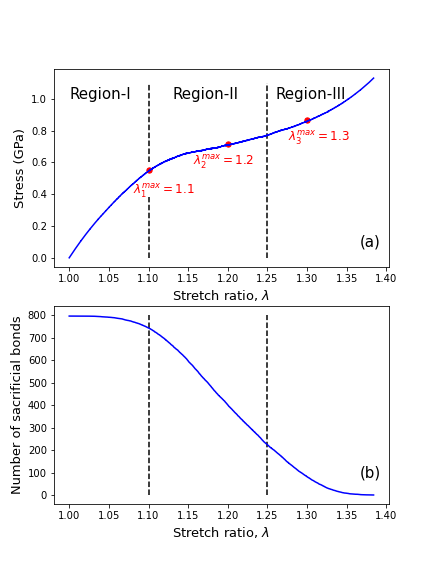}
	\caption{(a) The mean stress-strain response of the polymer under monotonic loading obtained using kinetic model. It shows  three distinct regions which are roughly demarcated   by the vertical dotted lines. $\lambda_1^{max}$, $\lambda_2^{max}$ and $\lambda_3^{max}$ correspond to the maximum strain applied in the three different cyclic loading protocols. (b) The mean number of sacrificial bonds for a given strain for monotonic loading.}
	\label{fig:4}
\end{figure} 

We start with a polymer of length $L_c$ with zero sacrificial bonds and equilibrate the system at zero force.  After equilibration, to do cyclic loading, we pull at a constant velocity such that $v=dx/dt$, where $x=\lambda L_a(t)$ is the end to end distance. The time-dependent stress  $\sigma(x, L_a(t))$ is calculated using Eq.~(\ref{eq:3}) and the corresponding rates are determined.  The polymer is pulled up to a pre-decided stretch ratio $\lambda^{max}$ after which the pulling velocity is reversed to $-v$, and the polymer is stretched back to zero force. This completes one loading cycle.
\begin{figure*}
	\graphicspath{{images/}}
	\centering
	\includegraphics[width=1.0 \linewidth]{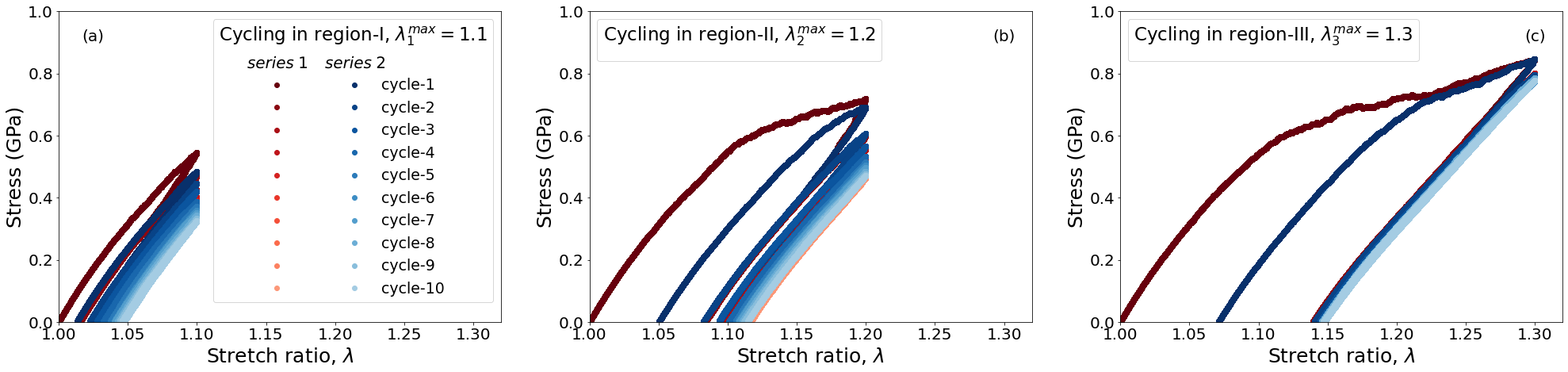}
	\caption{The macroscopic stress-stretch  response for cyclic loading for (a) $\lambda^{max}=1.1$ (region-I), (b)  $\lambda^{max}=1.2$ (region-II), and (c)  $\lambda^{max}=1.3$ (region-III). In all the three cases, the response shows moving hysteresis loops which saturate with loading cycles for both series 1 (first 10 cycles) and series 2 (next 10 cycles after 60 minutes relaxation at $F=0$) loading.}
	\label{fig:stressstrain}
\end{figure*}

\section{\label{Results} Results and Discussion}

\subsection{Uniaxial loading}

To establish the effectiveness of the proposed kinetic model, we first simulate response of the fibril chain polymer to monotonically increasing load. The average macroscopic response obtained from $16$ realizations is shown in Fig.~\ref{fig:4}(a). For each run, the system is first equilibrated at zero force, after which displacement (end to end distance) is increased at a constant velocity of $125\;nm/s$. The macroscopic response exhibits three distinct regions: an initial region (region-I) where stress increases linearly with strain, an intermediate region where stress is weakly increasing with strain (region-II) and a final region where the stress increases non-linearly with strain (region-III). These qualitative features, of three distinct regions, of the macroscopic response are consistent with what has been observed in  pulling experiments of collagen fibril~\cite{LIU2018cyclicload,SVENSSON20132476}.

The existence of three distinct regimes is better understood in terms of the number of the intact sacrificial bonds at any given strain. In Fig.~\ref{fig:4}(b), we show the mean number of sacrificial bonds for a given applied strain. For small strains, corresponding to region-I there is only a marginal decrease from its initial equilibrium value. Further increase in strain, corresponding to region-II, results in a sharp decrease in the number of sacrificial bonds, thereby releasing hidden lengths and causing relaxation in the stresses. Finally all sacrificial bonds are broken in region-III. The change in slope of the stress-strain curve in region-II occurs due to breaking of sacrificial bonds.

\subsection{Cyclic loading}

We next simulate the cyclic loading patterns reported in Ref.~\cite{LIU2018cyclicload} to compare the characteristic features of the mechanical response seen in the experiment with our simulations. Cyclic load is applied such that in each cycle the chain is stretched upto a maximum stretch ratio, $\lambda^{max}$. As in Ref.~\cite{LIU2018cyclicload}, we also consider $\lambda^{max}$ to lie in the three distinct regimes by choosing it to be 
$\lambda^{max}=1.1, 1.2, 1.3$ (the corresponding positions on the macroscopic response is shown by red circles in Fig.~\ref{fig:4}(a)) and these stretch ratios are representative points of regions I, II and III. The fibril is subjected to cyclic loading using the protocol described in Sec.~\ref{sec:protocol} with pulling speed $v= 125 \;nm/s$, chosen to be same as in experiment~\cite{LIU2018cyclicload}. The polymer is subjected to $10$ loading cycles (series 1)  and then relaxed at zero force for $60$ minutes, and then subjected to $10$ more loading cycles (series 2).
\begin{figure}
	\graphicspath{{images/}}
	\centering
	\includegraphics[width=1.0 \linewidth]{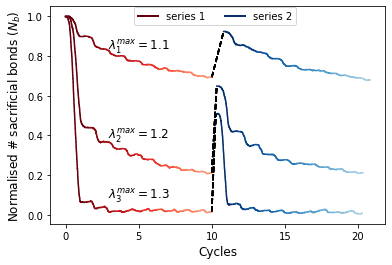}
	\caption{ The variation of number of sacrificial bonds with loading cycles for both series 1 and series 2 and for different stretch ratios, $\lambda^{max}$. The dashed line corresponds to discontinuity due to relaxation before series 2 loading. After relaxation, there is a partial recovery in number of sacrificial bonds for all $\lambda^{max}$. Color scheme used for cycles is same as in Fig.~\ref{fig:stressstrain} }
	\label{fig:8}
\end{figure}

We first present results for the variation of the stress-stretch      curve with cycles. The stress-stretch ratio  curves show hysteresis, as evident in Fig.~\ref{fig:stressstrain}. The first cycle exhibits hysteresis as well as residual strain at a completely unloaded state. Further cycling results in the subsequent hysteresis loops to shift to the right implying accumulation of residual strains. The hysteresis loops eventually tend to reach a steady state with number of cycles for both the series and for all three representative values of $\lambda^{max}$. These features from the  simulations of the  kinetic model are consistent with the observed trends in the experiment~\cite{LIU2018cyclicload}. 
\begin{figure}
	\graphicspath{{images/}}
	\centering
	\includegraphics[width=1.0 \linewidth]{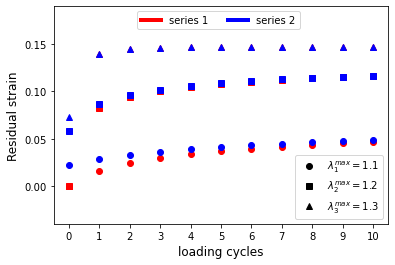}
	\caption{The evolution of residual strain with number of loading cycles for both series 1 and series 2 for different stretch ratios, $\lambda^{max}$. The magnitude and final saturated value of residual strain depends on the maximum stretch ratio, $\lambda^{max}$.}
	\label{fig:6}
\end{figure}
\begin{figure}
	\graphicspath{{images/}}
	\centering
	\includegraphics[width=1.0 \linewidth]{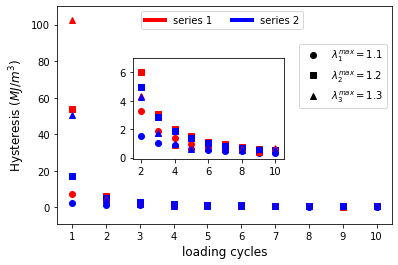}
	\caption{ The evolution of energy dissipation with number of loading cycles for both series 1 and series 2 and for different stretch ratios, $\lambda^{max}$. The data from 2nd cycle onward is zoomed and shown in the inset figure.}
	\label{fig:7}
\end{figure}

In the associated number of intact sacrificial bonds, shown in Fig.~\ref{fig:8} with fading shades of red and blue for series 1 and 2 respectively, the progressive breakage patterns with increasing cycles is clearly evident. For $\lambda^{max}=1.1$, the first cycle results in breakage of $10\%$ bonds and in subsequent 9 cycles  there is a further gradual reduction in sacrificial bonds, slowly reaching a steady state. During the waiting interval, bonds reform (shown with dashed line). The cyclic loading of series 2 causes the number of sacrificial bonds to gradually decrease again. For $\lambda_2^{max}$, however, most of the breakage occurs in the first cycle as the number of shows a dramatic decrease (by more than $50\%$). Subsequent cycles show comparatively lower rate of breakage per cycle. Interestingly, for a similar waiting interval, the reformation of bonds is significantly higher than for $\lambda_1^{max}$ and this could be attributed to the comparatively larger available length from more number of broken bonds. For the cyclic loads with $\lambda_3^{max}$, first cycle results in breakage of more than $90\%$ of the sacrificial bonds. Since most bonds are already broken further cycling does not affect the overall status of intact bonds appreciably. Waiting period recovers $50\%$ of the initial bonds which again break primarily in the first cycle of the series 2.

The residual strain accumulates with increasing cycles and reaches a steady state for both series 1 and series 2  loading for all three $\lambda^{max}$ (see Fig.~\ref{fig:6}). During the relaxation period between the two series, the residual strain reduces by approximately $50\%$. The magnitude of the residual strain when steady state is reached depends on  $\lambda^{max}$ (see Fig.~\ref{fig:6}). It can be seen that the steady state residual strains follows the order of $\lambda_3^{max} >\lambda_2^{max}>\lambda_1^{max}$, in agreement with the experiments~\cite{LIU2018cyclicload}. The residual strain increasing with $\lambda^{max}$ is due to the larger number of  sacrificial bonds breaking in the first cycle itself for higher $\lambda^{max}$, as shown in Fig.~\ref{fig:8}. It can also be seen that number of sacrificial bonds reform during relaxation which accounts for recovery in residual strain.
\begin{figure}
	\graphicspath{{images/}}
	\centering
	\includegraphics[width=1.0 \columnwidth]{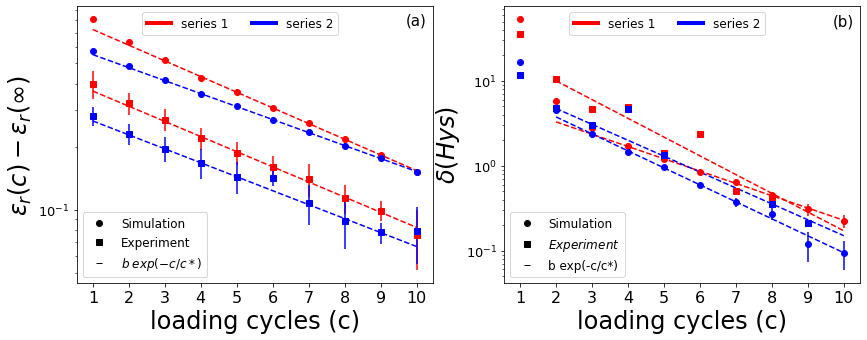}
	\caption{The variation of the deviation of  (a) residual strain and (b) area of hysteresis loop from their respective steady state values with number of loading cycles. The corresponding experimental data from Liu et. al. (2018)~\cite{LIU2018cyclicload} are shown with squares. Both quantities approaches steady state exponentially. The best fits are shown by  dashed lines.}
	\label{fig:9}
\end{figure}

The energy dissipated per cycle (area under the loading-unloading curve) decreases with increase in the number of cycles and reaches steady state for both series 1 and series 2 loading (see Fig.~\ref{fig:7}) for all chosen stretch ratios. There is a partial recovery in energy dissipation after relaxation as seen from first cycle of series 2 loading (see Figs.~\ref{fig:stressstrain} and \ref{fig:7}). The area of the hysteresis loop after the first cycle also follows the  pattern $\lambda_3^{max} >\lambda_2^{max}>\lambda_1^{max}$. This is because the first cycle of region-III has maximum number of sacrificial bond breaking compared to the other two regions, as evident from Fig.~\ref{fig:8}. Restoration of sacrificial bonds on relaxation accounts for recovery in energy dissipation. 
\begin{figure*}
	\graphicspath{{images/}}
	\centering
	\includegraphics[width=1.0 \linewidth]{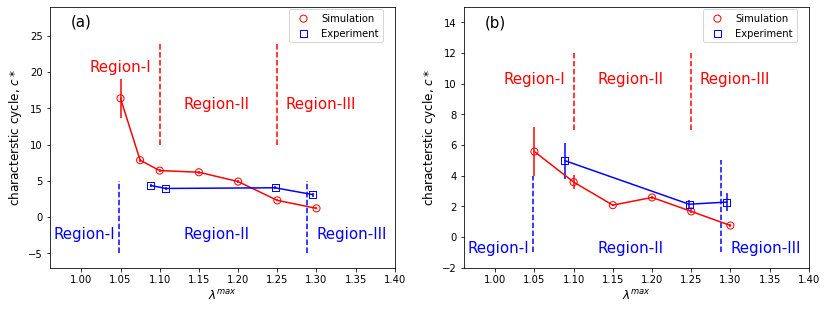}
	\caption{The variation of characteristic cycle $c^*$ with the maximum stretch ratio, $\lambda^{max}$ obtained for (a) residual strain and (b) energy dissipation.}
	\label{c_vs_lambda}
\end{figure*}

We now quantify the approach of residual strain and energy dissipation to their respective steady state values. We find that the deviation of residual strain and energy dissipation from their steady state value has an exponential decrease to zero with number of loading cycles (see Fig.~\ref{fig:9}, where the data for $\lambda^{max}=1.1$ and $\lambda^{max}=1.2$ are shown). We extract the experimental data for these quantities from Ref.~\cite{LIU2018cyclicload} and find that the exponential decrease is also seen in experiment (see Fig.~\ref{fig:9}). This allow us to determine a characteristic cycle number $c^*$ defined as:
\begin{equation}
	\epsilon_r(c)-\epsilon_r(\infty) \propto e^{-c/c^*},
\end{equation}
where $\epsilon_r(c)$ is residual strain at cycle $c$, $\epsilon_r(\infty$) is the steady state value of residual strain.

We compare the characteristic number of cycles, $c^*$, obtained for residual strain from simulations and experiments of Liu et al.~\cite{LIU2018cyclicload} in Fig.~\ref{c_vs_lambda}(a). We use the average $c^*$ of series 1 and 2 for the both simulations and experimental data. Since $c^*$ is not quoted in the experiments, we fit the extracted experimental data to obtain $c^*$.  From simulations, for small $\lambda^{max}$, in regime I, we find the polymeric chain takes larger number of cycles ($\approx16$)  to reach steady state. This large value of $c^*$ for small stretch ratio is understood as ideally, polymer should take infinite cycles to reach steady state within elastic regime. With increasing $\lambda^{max}$, $c^*$ decreases. In region II, the steady state is reached at significantly lower cycles ($\approx$5) and there is marginal decrease with increasing $\lambda^{max}$. Further increase in $\lambda^{max}$, corresponding to region III, shows again a further drop in $c^*$ implying faster approach to steady state in stress-lambda response. Experimental data compares very well in the region II as it also exhibits marginal change with increasing $\lambda^{max}$, and in region III there is  decrease in $c^*$ with increasing $\lambda^{max}$. 

We also compare the value of characteristic cycle $c^*$, obtained for energy dissipation from simulations and the extracted experimental data as shown in Fig.~\ref{c_vs_lambda}(b). We obtain a similar trend of $c^*$ with $\lambda^{max}$ for energy dissipation also. The value of $c^*$ is large in region-I, then it decreases with $\lambda^{max}$, it shows some plateau in region-II and then further decreases sharply in region-III. Again, we see a good match with experimental results.

\begin{figure}
	\graphicspath{{images/}}
	\centering
	\includegraphics[width=1.0 \linewidth]{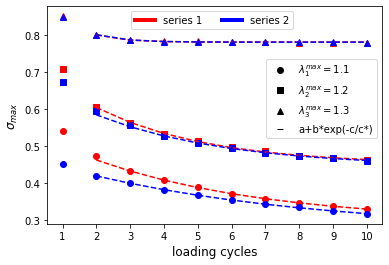}
	\caption{The variation of peak stress  with number of loading cycles for different $\lambda_1^{max}$, before (series 1) and after relaxation (series 2). Peak values of stress depends on $\lambda^{max}$ and  show a partial recovery on relaxation (series 2). }
	\label{fig:10}
\end{figure}

Finally, we study two more quantities studied in the experiment: peak stress and elastic modulus. The peak stress (stress at $\lambda^{max}$) decreases with number of cycles for both series 1 and 2 and for all three stretch ratios (see Fig.~\ref{fig:10}). It also approaches the steady state exponentially . 
The peak stress in the first cycle in a particular region depends on choice of $\lambda^{max}$ and follows the order: $\sigma( \lambda_3^{max}) >\sigma(\lambda_2^{max})>\sigma(\lambda_1^{max})$.

\begin{figure}
	\graphicspath{{images/}}
	\centering
	\includegraphics[width=1.0 \linewidth]{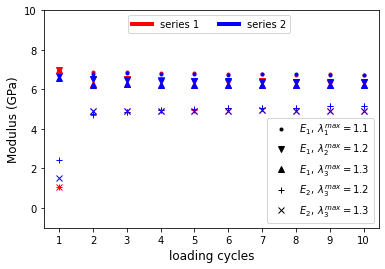}
	\caption{Elastic modulus $E_1$ remains invariant to cyclic loading for both series and in all three regions while $E_2$ (for region-II and III) becomes constant after first cycle and then remains invariant to cyclic loading.}
	\label{fig:12}
\end{figure}

We define two elastic moduli $E_1$ and $E_2$ in accordance with the experimental study~\cite{LIU2018cyclicload}.  The elastic modulus $E_1$ is calculated from the slope of the stress-stretch (Fig.~\ref{fig:stressstrain}) curve  up to $\lambda \approx 1.02$ whlie $E_2$ is calculated from the slope  (where $d \sigma / d \lambda \approx$ constant) of the stress-stretch curve in region-II. 
We find that $E_1$ is not affected by cyclic loading for all three stretch ratios while $E_2$ becomes a constant after the first loading cycle and then remains invariant to cyclic loading for $\lambda_2^{max}$ and $\lambda_3^{max}$ in both series loading (see Fig.~\ref{fig:12}). However, the extent of transition region between region-I to region-II decreases with cycles within $\lambda^{max}=1.2,\;1.3$ (see Fig.~\ref{fig:stressstrain}(b) and (c)) and these feature are also observed in the experiment. Our results are in good agreement with the cyclic loading experiment (see Figs.~3, 4, 5 in Ref.~\cite{LIU2018cyclicload}).

\section{\label{conclusion} Conclusions and discussion}

Experimentally, the stress-stretch response of a single collagen fibril subject to cyclic loading~\cite{SHEN2008cyclicload,LIU2018cyclicload} within a fixed stretch ratio $\lambda$ is known to show moving hysteresis loops and residual strains that increase and saturate with number of cycles. The fibril is known to show recovery in energy dissipation as well as residual strains on relaxation. These features were thought to be related to the presence of sacrificial bonds within the fibril~\cite{LIU2018cyclicload}. To test this hypothesis, we develop a stochastic kinetic model specifically for collagen fibril. The model treats the collagen fibril to be a polymeric chain that has hidden lengths secured by sacrificial bonds. The two primary ingredients of the model are: a reference stress-stretch relation for the available length of the polymer and stochastic formation and fragmentation of sacrificial bonds. The reference stress-stretch relation is first established from molecular dynamics simulations of an existing coarse-grain fibril model~\cite{MALASPINA2017549}. The kinetic model incorporates formation and breakage of sacrificial bonds and release of hidden lengths based on Bell's theory. We estimated the model parameters by comparing with available experimental data and used kinetic Monte Carlo methods to simulate the cyclic loading experiment.

The model qualitatively reproduces the main features of the experiment such as time evolution of hysteresis loops, energy dissipation, peak stress and residual strain etc. It is shown that these quantities approach their respective steady states exponentially with the number of loading cycles. We find that the characteristic cycle number associated with this exponential decay is in close agreement with the characteristic cycle number extracted from the reported  experimental data. The breaking of sacrificial bonds is responsible for hysteresis (energy dissipation) and the corresponding release of hidden lengths appears as residual strain. The magnitude of hysteresis, peak stress and residual strain after first cycle is proportional to maximum stretch ration $\lambda^{max}$. The recovery of the fibril is proportional to the relaxation time and spontaneous formation and breaking of sacrificial bonds at zero force is a possible healing mechanism in the collagen fibril.

The presence of a characteristic cycle number has significance in the description of the time dependent cyclic response of collagen. In particular, it has the potential of being utilised for comparison of fibril response across animals, ages, stages of disease, level of hierarchy, response to medication, etc. This is a promising area for future experimental investigation.

The kinetic model is able to reproduce the majority of the characteristic features of the fatigue experimental data in Ref.~\cite{LIU2018cyclicload}, thus providing an insight into the essential mechanisms at work. One feature that it is not able to explain is the experimentally observed increased  strength of the fibril post cyclic loading. This could be due to  permanent rearrangement of molecules inside the fibril, which a three-dimensional model of a fibril incorporating detailed microscopic interactions may be able to account for.  It would thus be of interest to develop a coarse grained model for the fibril that incorporates sacrificial bonds. In addition, it will provide a microscopic basis for the validity of the kinetic model, as well as allow for a determination of parameters. This is a promising area for future research.

Interestingly, at tissue scale also, the stress-strain response exhibits moving hysteresis loops, residual strain, etc~\cite{veres2013cross}. Linka et. al.~\cite{linka2016mechanics}, proposed a constitutive damage model that reproduces  the experimental results of the tendon overloading experiment~\cite{veres2013cross}. The kinetic model described in this paper, with suitable modifications, would also be ideally suited to explain the results at tissue level.

\begin{acknowledgments}
The simulations were carried out on the high performance computing machines Nandadevi at the Institute of Mathematical Sciences. We thank David C. Malaspina for helpful discussions in implementing the MD model.
\end{acknowledgments}

\appendix*
\section{\label{appendix:a} Coarse grained collagen fibril model}

In this appendix, we describe the coarse-grained model for the fibril that we have used for MD simulations. The model is from Ref.~\cite{DEPALLE20151,MALASPINA2017549}.

\begin{table}
	\centering
	\caption{\label{table:t2} The parameters for the  MD-model of fibril. }
	\begin{ruledtabular}
		\begin{tabular}{p{7.7cm} c}
			Model parameters & Value \\
			\hline
			$\epsilon$- LJ energy parameter ($kcal \;mol^{-1}$) & 6.87 \\
			$\sigma$- LJ distance parameter ($\AA $) & 14.72 \\
			$\theta_0$- Equilibrium bending angle ($degree$) & 180\\
			$k_{\theta}$- Bending strength constant ($kcal \;mol^{-1}\;rad^{-2} $) & 14.98 \\
			$r_0$- Equilibrium distance (tropocollagen) [$\AA $] & 14.00 \\
			$r_1$- critical hyperelastic distance (tropocollagen) [$\AA $] & 18.20  \\
			$r_{break}$- bond breaking distance (tropocollagen) & 21.00 \\
			$k_{T0}$- Stretching strength constant (tropocollagen) [$kcal \;mol^{-1}\; \AA^{-2} $] & 17.13 \\
			$k_{T1}$- Stretching strength constant (tropocollagen) [$kcal \;mol^{-1}\; \AA^{-2}$] & 97.66 \\
			$r_0$- Equilibrium distance (divalent crosslink) [$\AA $] & 10.00 \\
			$r_1$- critical hyperelastic distance (divalent crosslink) [$\AA $] & 12.00  \\
			$r_{break}$- bond breaking distance (divalent crosslink) & 14.68 \\
			$k_{T0}$- Stretching strength constant (divalent crosslink) [$kcal \;mol^{-1}\; \AA^{-2} $] & 0.20 \\
			$k_{T1}$- Stretching strength constant (divalent crosslink) [$kcal \;mol^{-1}\; \AA^{-2}$] & 41.84 \\
			$r_0$- Equilibrium distance (trivalent crosslink) [$\AA $] & 8.60 \\
			$r_1$- critical hyperelastic distance (trivalent crosslink) [$\AA $] & 12.20  \\
			$r_{break}$- Bond breaking distance (trivalent crosslink) & 14.89 \\
			$k_{T0}$- Stretching strength constant (trivalent crosslink) [$kcal \;mol^{-1}\; \AA^{-2} $] & 0.20 \\
			$k_{T1}$- Stretching strength constant (trivalent crosslink) [$kcal \;mol^{-1}\; \AA^{-2}$] & 54.60  \\
			$m$- mass of tropocollagen bead [$a.m.u$] & 1358.7 \\
		\end{tabular}
	\end{ruledtabular}
	\label{table:3}
\end{table}
 A collagen molecule is represented by $215$ beads connected with spring to each other. The distance between two consecutive bead is $b=1.4 \;nm$, which is roughly equals to the diameter of collagen molecule. Five collagen molecules are arranged in staggered manner in z-direction while in pentagonal geometry in x-y plane to form a microfibril. This staggered arrangement of collagen molecules give rise to the characteristic D-period of collagen fibril ($67 nm$). The diameter of a single microfibril is $\approx 3.5\; nm$. Terminal beads of each tropocollagen molecule forms a divalent or trivalent cross-link within a microfibril. In divalent cross-link, end beads of a molecule forms a single connection with a nearest bead from it's neighbouring molecule while in case of trivalent cross-link, the end beads forms two connection with the closest beads from it's nearest and next-nearest collagen molecule. These terminal connections represents the enzymatic cross-links in fibril. We have considered the case with hundred percent cross-link ($\beta=100\%$), which means all the terminal ends will from a cross-links with their neighbouring molecule. The ratio of trivalent ($33\%$) and divalent ($66\%$) cross-links has been kept fixed. Now, 37 of these microfibrils are arranged in a hexagonal close packing to represent a collagen fibril. The length and diameter of fibril model are $343.6\;nm$ and $25.9\;nm$ respectively. The periodic boundary conditions has been used to mimic the fibril of infinite length. Periodic boundary condition ensures the D-periodic pattern of the fibril structure.

The non bonded interaction between beads of  fibril is given by Lenard-Jones potential as :
\begin{equation}
U_{LJ}= 4 \epsilon \left[ \left( \frac{\sigma}{r} \right)^{12} - \left( \frac{\sigma}{r} \right)^{6} \right],
\end{equation}
where $r$ is the distance between interacting beads and $\sigma$ is the distance parameter and $\epsilon$ is energy parameter (depth of potential).

The bending energy ($U_{\theta}$) between three consecutive beads of collagen molecule is given by harmonic interaction as :
\begin{equation}
U_{\theta} = k_{\theta}(\theta-\theta_0)^2
\end{equation}
where $k_{\theta}$ is bending strength and $\theta_0$ is equilibrium angle.

The interaction between bonded beads is defined by a bi-harmonic potential as:
\begin{equation}
F_{bond}= \pdv{U_{bond}}{r} =  \begin{cases} 
k_{T0}(r-r_0) & \mathrm{for} \; r  < r_1, \\
k_{T1}(r-r_0)  & \mathrm{for} \; r_1\leq r < r_{break}, \\
0 &\mathrm{for} \; r> r_{break}. 
\end{cases}
\end{equation}
where $r_0$ is the equilibrium distance between two beads, $k_{T0}$ and $k_{T1}$ are spring constants between distances $0$ to $r_1$ and $0$ to $r_{break}$. 

The simulations were performed using LAMMPS~\cite{LAMMPS}. Time step was set to $\Delta t=10\;fs$, and the equations of motion were integrated with langevin thermostat with drag coefficient $1000 \;fs$ and temperature 310 K. The fibril was equilibrated for $20\; ns$ and then a constant strain rate of $10^7 \;s^{-1}$ was applied.  All the parameters used in simulation are given in Table~\ref{table:t2}. These parameters have been developed for collagen molecules in Refs.~\cite{Buehler2006a,Buehler12006b} and specifically for the fibril model in Refs.~\cite{DEPALLE20151,MALASPINA2017549}.


\begin{thebibliography}{41}%
	\makeatletter
	\providecommand \@ifxundefined [1]{%
		\@ifx{#1\undefined}
	}%
	\providecommand \@ifnum [1]{%
		\ifnum #1\expandafter \@firstoftwo
		\else \expandafter \@secondoftwo
		\fi
	}%
	\providecommand \@ifx [1]{%
		\ifx #1\expandafter \@firstoftwo
		\else \expandafter \@secondoftwo
		\fi
	}%
	\providecommand \natexlab [1]{#1}%
	\providecommand \enquote  [1]{``#1''}%
	\providecommand \bibnamefont  [1]{#1}%
	\providecommand \bibfnamefont [1]{#1}%
	\providecommand \citenamefont [1]{#1}%
	\providecommand \href@noop [0]{\@secondoftwo}%
	\providecommand \href [0]{\begingroup \@sanitize@url \@href}%
	\providecommand \@href[1]{\@@startlink{#1}\@@href}%
	\providecommand \@@href[1]{\endgroup#1\@@endlink}%
	\providecommand \@sanitize@url [0]{\catcode `\\12\catcode `\$12\catcode
		`\&12\catcode `\#12\catcode `\^12\catcode `\_12\catcode `\%12\relax}%
	\providecommand \@@startlink[1]{}%
	\providecommand \@@endlink[0]{}%
	\providecommand \url  [0]{\begingroup\@sanitize@url \@url }%
	\providecommand \@url [1]{\endgroup\@href {#1}{\urlprefix }}%
	\providecommand \urlprefix  [0]{URL }%
	\providecommand \Eprint [0]{\href }%
	\providecommand \doibase [0]{https://doi.org/}%
	\providecommand \selectlanguage [0]{\@gobble}%
	\providecommand \bibinfo  [0]{\@secondoftwo}%
	\providecommand \bibfield  [0]{\@secondoftwo}%
	\providecommand \translation [1]{[#1]}%
	\providecommand \BibitemOpen [0]{}%
	\providecommand \bibitemStop [0]{}%
	\providecommand \bibitemNoStop [0]{.\EOS\space}%
	\providecommand \EOS [0]{\spacefactor3000\relax}%
	\providecommand \BibitemShut  [1]{\csname bibitem#1\endcsname}%
	\let\auto@bib@innerbib\@empty
	\bibitem [{\citenamefont {Kadler}\ \emph {et~al.}(2007)\citenamefont {Kadler},
		\citenamefont {Baldock}, \citenamefont {Bella},\ and\ \citenamefont
		{Boot-Handford}}]{Kalder2007}%
	\BibitemOpen
	\bibfield  {author} {\bibinfo {author} {\bibfnamefont {K.~E.}\ \bibnamefont
			{Kadler}}, \bibinfo {author} {\bibfnamefont {C.}~\bibnamefont {Baldock}},
		\bibinfo {author} {\bibfnamefont {J.}~\bibnamefont {Bella}},\ and\ \bibinfo
		{author} {\bibfnamefont {R.~P.}\ \bibnamefont {Boot-Handford}},\ }\bibfield
	{title} {\bibinfo {title} {Collagens at a glance},\ }\href
	{https://doi.org/10.1242/jcs.03453} {\bibfield  {journal} {\bibinfo
			{journal} {Journal of Cell Science}\ }\textbf {\bibinfo {volume} {120}},\
		\bibinfo {pages} {1955} (\bibinfo {year} {2007})}\BibitemShut {NoStop}%
	\bibitem [{\citenamefont {Fratzl}(2008)}]{Fratzl2008}%
	\BibitemOpen
	\bibfield  {author} {\bibinfo {author} {\bibfnamefont {P.}~\bibnamefont
			{Fratzl}},\ }\bibfield  {title} {\bibinfo {title} {Collagen: Structure and
			mechanics, an introduction.},\ }\href
	{https://doi.org/10.1007/978-0-387-73906-9_1} {\bibfield  {journal} {\bibinfo
			{journal} {Springer:New York}\ } (\bibinfo {year} {2008})}\BibitemShut
	{NoStop}%
	\bibitem [{\citenamefont {Shoulders}\ and\ \citenamefont
		{Raines}(2009)}]{Shoulders2009}%
	\BibitemOpen
	\bibfield  {author} {\bibinfo {author} {\bibfnamefont {M.~D.}\ \bibnamefont
			{Shoulders}}\ and\ \bibinfo {author} {\bibfnamefont {R.~T.}\ \bibnamefont
			{Raines}},\ }\bibfield  {title} {\bibinfo {title} {Collagen structure and
			stability},\ }\href@noop {} {\bibfield  {journal} {\bibinfo  {journal}
			{Annual review of biochemistry}\ }\textbf {\bibinfo {volume} {78}},\ \bibinfo
		{pages} {929} (\bibinfo {year} {2009})}\BibitemShut {NoStop}%
	\bibitem [{\citenamefont {Ottani}\ \emph {et~al.}(2002)\citenamefont {Ottani},
		\citenamefont {Martini}, \citenamefont {Franchi}, \citenamefont {Ruggeri},\
		and\ \citenamefont {Raspanti}}]{OTTANI2002}%
	\BibitemOpen
	\bibfield  {author} {\bibinfo {author} {\bibfnamefont {V.}~\bibnamefont
			{Ottani}}, \bibinfo {author} {\bibfnamefont {D.}~\bibnamefont {Martini}},
		\bibinfo {author} {\bibfnamefont {M.}~\bibnamefont {Franchi}}, \bibinfo
		{author} {\bibfnamefont {A.}~\bibnamefont {Ruggeri}},\ and\ \bibinfo {author}
		{\bibfnamefont {M.}~\bibnamefont {Raspanti}},\ }\bibfield  {title} {\bibinfo
		{title} {Hierarchical structures in fibrillar collagens},\ }\href
	{https://doi.org/https://doi.org/10.1016/S0968-4328(02)00033-1} {\bibfield
		{journal} {\bibinfo  {journal} {Micron}\ }\textbf {\bibinfo {volume} {33}},\
		\bibinfo {pages} {587} (\bibinfo {year} {2002})}\BibitemShut {NoStop}%
	\bibitem [{\citenamefont {Fratzl}\ and\ \citenamefont
		{Weinkamer}(2007)}]{FRATZL20071263}%
	\BibitemOpen
	\bibfield  {author} {\bibinfo {author} {\bibfnamefont {P.}~\bibnamefont
			{Fratzl}}\ and\ \bibinfo {author} {\bibfnamefont {R.}~\bibnamefont
			{Weinkamer}},\ }\bibfield  {title} {\bibinfo {title} {Nature’s hierarchical
			materials},\ }\href
	{https://doi.org/https://doi.org/10.1016/j.pmatsci.2007.06.001} {\bibfield
		{journal} {\bibinfo  {journal} {Progress in Materials Science}\ }\textbf
		{\bibinfo {volume} {52}},\ \bibinfo {pages} {1263} (\bibinfo {year}
		{2007})}\BibitemShut {NoStop}%
	\bibitem [{\citenamefont {Parenteau-Bareil}\ \emph {et~al.}(2010)\citenamefont
		{Parenteau-Bareil}, \citenamefont {Gauvin},\ and\ \citenamefont
		{Berthod}}]{parenteau2010collagen}%
	\BibitemOpen
	\bibfield  {author} {\bibinfo {author} {\bibfnamefont {R.}~\bibnamefont
			{Parenteau-Bareil}}, \bibinfo {author} {\bibfnamefont {R.}~\bibnamefont
			{Gauvin}},\ and\ \bibinfo {author} {\bibfnamefont {F.}~\bibnamefont
			{Berthod}},\ }\bibfield  {title} {\bibinfo {title} {Collagen-based
			biomaterials for tissue engineering applications},\ }\href@noop {} {\bibfield
		{journal} {\bibinfo  {journal} {Materials}\ }\textbf {\bibinfo {volume}
			{3}},\ \bibinfo {pages} {1863} (\bibinfo {year} {2010})}\BibitemShut
	{NoStop}%
	\bibitem [{\citenamefont {Lee}\ \emph {et~al.}(2001)\citenamefont {Lee},
		\citenamefont {Singla},\ and\ \citenamefont {Lee}}]{lee2001biomedical}%
	\BibitemOpen
	\bibfield  {author} {\bibinfo {author} {\bibfnamefont {C.~H.}\ \bibnamefont
			{Lee}}, \bibinfo {author} {\bibfnamefont {A.}~\bibnamefont {Singla}},\ and\
		\bibinfo {author} {\bibfnamefont {Y.}~\bibnamefont {Lee}},\ }\bibfield
	{title} {\bibinfo {title} {Biomedical applications of collagen},\ }\href@noop
	{} {\bibfield  {journal} {\bibinfo  {journal} {International journal of
				pharmaceutics}\ }\textbf {\bibinfo {volume} {221}},\ \bibinfo {pages} {1}
		(\bibinfo {year} {2001})}\BibitemShut {NoStop}%
	\bibitem [{\citenamefont {Petruska}\ and\ \citenamefont
		{Hodge}(1964)}]{PETRUSKA1964}%
	\BibitemOpen
	\bibfield  {author} {\bibinfo {author} {\bibfnamefont {J.~A.}\ \bibnamefont
			{Petruska}}\ and\ \bibinfo {author} {\bibfnamefont {A.~J.}\ \bibnamefont
			{Hodge}},\ }\bibfield  {title} {\bibinfo {title} {A subunit model for the
			tropocollagen macromolecule},\ }\href@noop {} {\bibfield  {journal} {\bibinfo
			{journal} {Proceedings of the National Academy of Sciences of the United
				States of America}\ }\textbf {\bibinfo {volume} {51}},\ \bibinfo {pages}
		{871} (\bibinfo {year} {1964})}\BibitemShut {NoStop}%
	\bibitem [{\citenamefont {Orgel}\ \emph {et~al.}(2006)\citenamefont {Orgel},
		\citenamefont {Irving}, \citenamefont {Miller},\ and\ \citenamefont
		{Wess}}]{Orgel2006}%
	\BibitemOpen
	\bibfield  {author} {\bibinfo {author} {\bibfnamefont {J.~P.}\ \bibnamefont
			{Orgel}}, \bibinfo {author} {\bibfnamefont {T.~C.}\ \bibnamefont {Irving}},
		\bibinfo {author} {\bibfnamefont {A.}~\bibnamefont {Miller}},\ and\ \bibinfo
		{author} {\bibfnamefont {T.~J.}\ \bibnamefont {Wess}},\ }\bibfield  {title}
	{\bibinfo {title} {Microfibrillar structure of type I collagen in situ},\
	}\href@noop {} {\bibfield  {journal} {\bibinfo  {journal} {Proceedings of the
				National Academy of Sciences}\ }\textbf {\bibinfo {volume} {103}},\ \bibinfo
		{pages} {9001} (\bibinfo {year} {2006})}\BibitemShut {NoStop}%
	\bibitem [{\citenamefont {Light}\ and\ \citenamefont
		{Bailey}(1980)}]{Light1980}%
	\BibitemOpen
	\bibfield  {author} {\bibinfo {author} {\bibfnamefont {N.}~\bibnamefont
			{Light}}\ and\ \bibinfo {author} {\bibfnamefont {A.}~\bibnamefont {Bailey}},\
	}\bibfield  {title} {\bibinfo {title} {The chemistry of the collagen
			cross-links. purification and characterization of cross-linked polymeric
			peptide material from mature collagen containing unknown amino acids},\
	}\href@noop {} {\bibfield  {journal} {\bibinfo  {journal} {Biochemical
				Journal}\ }\textbf {\bibinfo {volume} {185}},\ \bibinfo {pages} {373}
		(\bibinfo {year} {1980})}\BibitemShut {NoStop}%
	\bibitem [{\citenamefont {Knott}\ and\ \citenamefont
		{Bailey}(1998)}]{KNOTT1998}%
	\BibitemOpen
	\bibfield  {author} {\bibinfo {author} {\bibfnamefont {L.}~\bibnamefont
			{Knott}}\ and\ \bibinfo {author} {\bibfnamefont {A.}~\bibnamefont {Bailey}},\
	}\bibfield  {title} {\bibinfo {title} {Collagen cross-links in mineralizing
			tissues: A review of their chemistry, function, and clinical relevance},\
	}\href {https://doi.org/https://doi.org/10.1016/S8756-3282(97)00279-2}
	{\bibfield  {journal} {\bibinfo  {journal} {Bone}\ }\textbf {\bibinfo
			{volume} {22}},\ \bibinfo {pages} {181} (\bibinfo {year} {1998})}\BibitemShut
	{NoStop}%
	\bibitem [{\citenamefont {Reiser}\ \emph {et~al.}(1992)\citenamefont {Reiser},
		\citenamefont {McCormick},\ and\ \citenamefont {Rucker}}]{Reiser1992}%
	\BibitemOpen
	\bibfield  {author} {\bibinfo {author} {\bibfnamefont {K.}~\bibnamefont
			{Reiser}}, \bibinfo {author} {\bibfnamefont {R.~J.}\ \bibnamefont
			{McCormick}},\ and\ \bibinfo {author} {\bibfnamefont {R.~B.}\ \bibnamefont
			{Rucker}},\ }\bibfield  {title} {\bibinfo {title} {Enzymatic and nonenzymatic
			cross-linking of collagen and elastin},\ }\href@noop {} {\bibfield  {journal}
		{\bibinfo  {journal} {FASEB J}\ }\textbf {\bibinfo {volume} {6}},\ \bibinfo
		{pages} {2439} (\bibinfo {year} {1992})}\BibitemShut {NoStop}%
	\bibitem [{\citenamefont {Thompson}\ \emph {et~al.}(2001)\citenamefont
		{Thompson}, \citenamefont {Kindt}, \citenamefont {Drake}, \citenamefont
		{Hansma}, \citenamefont {Morse},\ and\ \citenamefont
		{Hansma}}]{Thompson2001}%
	\BibitemOpen
	\bibfield  {author} {\bibinfo {author} {\bibfnamefont {J.~B.}\ \bibnamefont
			{Thompson}}, \bibinfo {author} {\bibfnamefont {J.~H.}\ \bibnamefont {Kindt}},
		\bibinfo {author} {\bibfnamefont {B.}~\bibnamefont {Drake}}, \bibinfo
		{author} {\bibfnamefont {H.~G.}\ \bibnamefont {Hansma}}, \bibinfo {author}
		{\bibfnamefont {D.~E.}\ \bibnamefont {Morse}},\ and\ \bibinfo {author}
		{\bibfnamefont {P.~K.}\ \bibnamefont {Hansma}},\ }\bibfield  {title}
	{\bibinfo {title} {Bone indentation recovery time correlates with bond
			reforming time},\ }\href {https://doi.org/10.1038/414773a} {\bibfield
		{journal} {\bibinfo  {journal} {Nature}\ }\textbf {\bibinfo {volume} {414}},\
		\bibinfo {pages} {773} (\bibinfo {year} {2001})}\BibitemShut {NoStop}%
	\bibitem [{\citenamefont {Sun}\ \emph {et~al.}(2002)\citenamefont {Sun},
		\citenamefont {Luo}, \citenamefont {Fertala},\ and\ \citenamefont
		{An}}]{SUN2002}%
	\BibitemOpen
	\bibfield  {author} {\bibinfo {author} {\bibfnamefont {Y.-L.}\ \bibnamefont
			{Sun}}, \bibinfo {author} {\bibfnamefont {Z.-P.}\ \bibnamefont {Luo}},
		\bibinfo {author} {\bibfnamefont {A.}~\bibnamefont {Fertala}},\ and\ \bibinfo
		{author} {\bibfnamefont {K.-N.}\ \bibnamefont {An}},\ }\bibfield  {title}
	{\bibinfo {title} {Direct quantification of the flexibility of type I
			collagen monomer},\ }\href@noop {} {\bibfield  {journal} {\bibinfo  {journal}
			{Biochemical and biophysical research communications}\ }\textbf {\bibinfo
			{volume} {295}},\ \bibinfo {pages} {382} (\bibinfo {year}
		{2002})}\BibitemShut {NoStop}%
	\bibitem [{\citenamefont {Sun}\ \emph {et~al.}(2004)\citenamefont {Sun},
		\citenamefont {Luo}, \citenamefont {Fertala},\ and\ \citenamefont
		{An}}]{SUN2004}%
	\BibitemOpen
	\bibfield  {author} {\bibinfo {author} {\bibfnamefont {Y.-L.}\ \bibnamefont
			{Sun}}, \bibinfo {author} {\bibfnamefont {Z.-P.}\ \bibnamefont {Luo}},
		\bibinfo {author} {\bibfnamefont {A.}~\bibnamefont {Fertala}},\ and\ \bibinfo
		{author} {\bibfnamefont {K.-N.}\ \bibnamefont {An}},\ }\bibfield  {title}
	{\bibinfo {title} {Stretching type II collagen with optical tweezers},\
	}\href@noop {} {\bibfield  {journal} {\bibinfo  {journal} {Journal of
				biomechanics}\ }\textbf {\bibinfo {volume} {37}},\ \bibinfo {pages} {1665}
		(\bibinfo {year} {2004})}\BibitemShut {NoStop}%
	\bibitem [{\citenamefont {Bozec}\ and\ \citenamefont
		{Horton}(2005)}]{BOZEC2005}%
	\BibitemOpen
	\bibfield  {author} {\bibinfo {author} {\bibfnamefont {L.}~\bibnamefont
			{Bozec}}\ and\ \bibinfo {author} {\bibfnamefont {M.}~\bibnamefont {Horton}},\
	}\bibfield  {title} {\bibinfo {title} {Topography and mechanical properties
			of single molecules of type I collagen using atomic force microscopy},\
	}\href {https://doi.org/https://doi.org/10.1529/biophysj.104.055228}
	{\bibfield  {journal} {\bibinfo  {journal} {Biophysical Journal}\ }\textbf
		{\bibinfo {volume} {88}},\ \bibinfo {pages} {4223} (\bibinfo {year}
		{2005})}\BibitemShut {NoStop}%
	\bibitem [{\citenamefont {Smith}\ \emph {et~al.}(1999)\citenamefont {Smith},
		\citenamefont {Sch{\"a}ffer}, \citenamefont {Viani}, \citenamefont
		{Thompson}, \citenamefont {Frederick}, \citenamefont {Kindt}, \citenamefont
		{Belcher}, \citenamefont {Stucky}, \citenamefont {Morse},\ and\ \citenamefont
		{Hansma}}]{Smith1999}%
	\BibitemOpen
	\bibfield  {author} {\bibinfo {author} {\bibfnamefont {B.~L.}\ \bibnamefont
			{Smith}}, \bibinfo {author} {\bibfnamefont {T.~E.}\ \bibnamefont
			{Sch{\"a}ffer}}, \bibinfo {author} {\bibfnamefont {M.}~\bibnamefont {Viani}},
		\bibinfo {author} {\bibfnamefont {J.~B.}\ \bibnamefont {Thompson}}, \bibinfo
		{author} {\bibfnamefont {N.~A.}\ \bibnamefont {Frederick}}, \bibinfo {author}
		{\bibfnamefont {J.}~\bibnamefont {Kindt}}, \bibinfo {author} {\bibfnamefont
			{A.}~\bibnamefont {Belcher}}, \bibinfo {author} {\bibfnamefont {G.~D.}\
			\bibnamefont {Stucky}}, \bibinfo {author} {\bibfnamefont {D.~E.}\
			\bibnamefont {Morse}},\ and\ \bibinfo {author} {\bibfnamefont {P.~K.}\
			\bibnamefont {Hansma}},\ }\bibfield  {title} {\bibinfo {title} {Molecular
			mechanistic origin of the toughness of natural adhesives, fibres and
			composites},\ }\href {https://doi.org/10.1038/21607} {\bibfield  {journal}
		{\bibinfo  {journal} {Nature}\ }\textbf {\bibinfo {volume} {399}},\ \bibinfo
		{pages} {761} (\bibinfo {year} {1999})}\BibitemShut {NoStop}%
	\bibitem [{\citenamefont {Rief}\ \emph {et~al.}(1997)\citenamefont {Rief},
		\citenamefont {Gautel}, \citenamefont {Oesterhelt}, \citenamefont
		{Fernandez},\ and\ \citenamefont {Gaub}}]{Reif1997}%
	\BibitemOpen
	\bibfield  {author} {\bibinfo {author} {\bibfnamefont {M.}~\bibnamefont
			{Rief}}, \bibinfo {author} {\bibfnamefont {M.}~\bibnamefont {Gautel}},
		\bibinfo {author} {\bibfnamefont {F.}~\bibnamefont {Oesterhelt}}, \bibinfo
		{author} {\bibfnamefont {J.~M.}\ \bibnamefont {Fernandez}},\ and\ \bibinfo
		{author} {\bibfnamefont {H.~E.}\ \bibnamefont {Gaub}},\ }\bibfield  {title}
	{\bibinfo {title} {Reversible unfolding of individual titin immunoglobulin
			domains by AFM},\ }\href {https://doi.org/10.1126/science.276.5315.1109}
	{\bibfield  {journal} {\bibinfo  {journal} {Science}\ }\textbf {\bibinfo
			{volume} {276}},\ \bibinfo {pages} {1109} (\bibinfo {year}
		{1997})}\BibitemShut {NoStop}%
	\bibitem [{\citenamefont {Orgel}\ \emph {et~al.}(2000)\citenamefont {Orgel},
		\citenamefont {Wess},\ and\ \citenamefont {Miller}}]{ORGEL2001}%
	\BibitemOpen
	\bibfield  {author} {\bibinfo {author} {\bibfnamefont {J.~P.}\ \bibnamefont
			{Orgel}}, \bibinfo {author} {\bibfnamefont {T.~J.}\ \bibnamefont {Wess}},\
		and\ \bibinfo {author} {\bibfnamefont {A.}~\bibnamefont {Miller}},\
	}\bibfield  {title} {\bibinfo {title} {The in situ conformation and axial
			location of the intermolecular cross-linked non-helical telopeptides of type
			I collagen},\ }\href@noop {} {\bibfield  {journal} {\bibinfo  {journal}
			{Structure}\ }\textbf {\bibinfo {volume} {8}},\ \bibinfo {pages} {137}
		(\bibinfo {year} {2000})}\BibitemShut {NoStop}%
	\bibitem [{\citenamefont {Svensson}\ \emph {et~al.}(2013)\citenamefont
		{Svensson}, \citenamefont {Mulder}, \citenamefont {Kovanen},\ and\
		\citenamefont {Magnusson}}]{SVENSSON20132476}%
	\BibitemOpen
	\bibfield  {author} {\bibinfo {author} {\bibfnamefont {R.}~\bibnamefont
			{Svensson}}, \bibinfo {author} {\bibfnamefont {H.}~\bibnamefont {Mulder}},
		\bibinfo {author} {\bibfnamefont {V.}~\bibnamefont {Kovanen}},\ and\ \bibinfo
		{author} {\bibfnamefont {S.}~\bibnamefont {Magnusson}},\ }\bibfield  {title}
	{\bibinfo {title} {Fracture mechanics of collagen fibrils: Influence of
			natural cross-links},\ }\href
	{https://doi.org/https://doi.org/10.1016/j.bpj.2013.04.033} {\bibfield
		{journal} {\bibinfo  {journal} {Biophysical Journal}\ }\textbf {\bibinfo
			{volume} {104}},\ \bibinfo {pages} {2476} (\bibinfo {year}
		{2013})}\BibitemShut {NoStop}%
	\bibitem [{\citenamefont {Liu}\ \emph {et~al.}(2010)\citenamefont {Liu},
		\citenamefont {Dodge}, \citenamefont {Kahn}, \citenamefont {Ballarini},\ and\
		\citenamefont {Eppell}}]{SHEN2010}%
	\BibitemOpen
	\bibfield  {author} {\bibinfo {author} {\bibfnamefont {Z.}~\bibnamefont
			{Liu}}, \bibinfo {author} {\bibfnamefont {M.~R.}\ \bibnamefont {Dodge}},
		\bibinfo {author} {\bibfnamefont {H.}~\bibnamefont {Kahn}}, \bibinfo {author}
		{\bibfnamefont {R.}~\bibnamefont {Ballarini}},\ and\ \bibinfo {author}
		{\bibfnamefont {S.~J.}\ \bibnamefont {Eppell}},\ }\bibfield  {title}
	{\bibinfo {title} {In vitro fracture testing of submicron diameter collagen
			fibril specimens},\ }\href
	{https://doi.org/https://doi.org/10.1016/j.bpj.2010.07.021} {\bibfield
		{journal} {\bibinfo  {journal} {Biophysical Journal}\ }\textbf {\bibinfo
			{volume} {99}},\ \bibinfo {pages} {1986} (\bibinfo {year}
		{2010})}\BibitemShut {NoStop}%
	\bibitem [{\citenamefont {Lorenzo}\ and\ \citenamefont
		{Caffarena}(2005)}]{LORENZO2005}%
	\BibitemOpen
	\bibfield  {author} {\bibinfo {author} {\bibfnamefont {A.~C.}\ \bibnamefont
			{Lorenzo}}\ and\ \bibinfo {author} {\bibfnamefont {E.~R.}\ \bibnamefont
			{Caffarena}},\ }\bibfield  {title} {\bibinfo {title} {Elastic properties,
			young's modulus determination and structural stability of the tropocollagen
			molecule: a computational study by steered molecular dynamics},\ }\href
	{https://doi.org/https://doi.org/10.1016/j.jbiomech.2004.07.011} {\bibfield
		{journal} {\bibinfo  {journal} {Journal of Biomechanics}\ }\textbf {\bibinfo
			{volume} {38}},\ \bibinfo {pages} {1527} (\bibinfo {year}
		{2005})}\BibitemShut {NoStop}%
	\bibitem [{\citenamefont {Buehler}(2006{\natexlab{a}})}]{Buehler2006a}%
	\BibitemOpen
	\bibfield  {author} {\bibinfo {author} {\bibfnamefont {M.~J.}\ \bibnamefont
			{Buehler}},\ }\bibfield  {title} {\bibinfo {title} {Atomistic and continuum
			modeling of mechanical properties of collagen: Elasticity, fracture, and
			self-assembly},\ }\href {https://doi.org/10.1557/jmr.2006.0236} {\bibfield
		{journal} {\bibinfo  {journal} {Journal of Materials Research}\ }\textbf
		{\bibinfo {volume} {21}},\ \bibinfo {pages} {1947} (\bibinfo {year}
		{2006}{\natexlab{a}})}\BibitemShut {NoStop}%
	\bibitem [{\citenamefont {Buehler}(2006{\natexlab{b}})}]{Buehler12006b}%
	\BibitemOpen
	\bibfield  {author} {\bibinfo {author} {\bibfnamefont {M.~J.}\ \bibnamefont
			{Buehler}},\ }\bibfield  {title} {\bibinfo {title} {Nature designs tough
			collagen: explaining the nanostructure of collagen fibrils},\ }\href@noop {}
	{\bibfield  {journal} {\bibinfo  {journal} {Proceedings of the National
				Academy of Sciences}\ }\textbf {\bibinfo {volume} {103}},\ \bibinfo {pages}
		{12285} (\bibinfo {year} {2006}{\natexlab{b}})}\BibitemShut {NoStop}%
	\bibitem [{\citenamefont {Gautieri}\ \emph {et~al.}(2009)\citenamefont
		{Gautieri}, \citenamefont {Buehler},\ and\ \citenamefont
		{Redaelli}}]{GAUTIERI2009}%
	\BibitemOpen
	\bibfield  {author} {\bibinfo {author} {\bibfnamefont {A.}~\bibnamefont
			{Gautieri}}, \bibinfo {author} {\bibfnamefont {M.~J.}\ \bibnamefont
			{Buehler}},\ and\ \bibinfo {author} {\bibfnamefont {A.}~\bibnamefont
			{Redaelli}},\ }\bibfield  {title} {\bibinfo {title} {Deformation rate
			controls elasticity and unfolding pathway of single tropocollagen
			molecules},\ }\href@noop {} {\bibfield  {journal} {\bibinfo  {journal}
			{Journal of the Mechanical Behavior of Biomedical Materials}\ }\textbf
		{\bibinfo {volume} {2}},\ \bibinfo {pages} {130} (\bibinfo {year}
		{2009})}\BibitemShut {NoStop}%
	\bibitem [{\citenamefont {Buehler}\ and\ \citenamefont
		{Wong}(2007)}]{BuehlerWong2007}%
	\BibitemOpen
	\bibfield  {author} {\bibinfo {author} {\bibfnamefont {M.~J.}\ \bibnamefont
			{Buehler}}\ and\ \bibinfo {author} {\bibfnamefont {S.~Y.}\ \bibnamefont
			{Wong}},\ }\bibfield  {title} {\bibinfo {title} {Entropic elasticity controls
			nanomechanics of single tropocollagen molecules},\ }\href
	{https://doi.org/https://doi.org/10.1529/biophysj.106.102616} {\bibfield
		{journal} {\bibinfo  {journal} {Biophysical Journal}\ }\textbf {\bibinfo
			{volume} {93}},\ \bibinfo {pages} {37} (\bibinfo {year} {2007})}\BibitemShut
	{NoStop}%
	\bibitem [{\citenamefont {Uzel}\ and\ \citenamefont
		{Buehler}(2011)}]{UZEL2011}%
	\BibitemOpen
	\bibfield  {author} {\bibinfo {author} {\bibfnamefont {S.~G.}\ \bibnamefont
			{Uzel}}\ and\ \bibinfo {author} {\bibfnamefont {M.~J.}\ \bibnamefont
			{Buehler}},\ }\bibfield  {title} {\bibinfo {title} {Molecular structure,
			mechanical behavior and failure mechanism of the c-terminal cross-link domain
			in type I collagen},\ }\href@noop {} {\bibfield  {journal} {\bibinfo
			{journal} {Journal of the Mechanical Behavior of Biomedical Materials}\
		}\textbf {\bibinfo {volume} {4}},\ \bibinfo {pages} {153} (\bibinfo {year}
		{2011})}\BibitemShut {NoStop}%
	\bibitem [{\citenamefont {Buehler}(2008)}]{BUEHLER2007}%
	\BibitemOpen
	\bibfield  {author} {\bibinfo {author} {\bibfnamefont {M.~J.}\ \bibnamefont
			{Buehler}},\ }\bibfield  {title} {\bibinfo {title} {Nanomechanics of collagen
			fibrils under varying cross-link densities: Atomistic and continuum
			studies},\ }\href
	{https://doi.org/https://doi.org/10.1016/j.jmbbm.2007.04.001} {\bibfield
		{journal} {\bibinfo  {journal} {Journal of the Mechanical Behavior of
				Biomedical Materials}\ }\textbf {\bibinfo {volume} {1}},\ \bibinfo {pages}
		{59} (\bibinfo {year} {2008})}\BibitemShut {NoStop}%
	\bibitem [{\citenamefont {Depalle}\ \emph {et~al.}(2015)\citenamefont
		{Depalle}, \citenamefont {Qin}, \citenamefont {Shefelbine},\ and\
		\citenamefont {Buehler}}]{DEPALLE20151}%
	\BibitemOpen
	\bibfield  {author} {\bibinfo {author} {\bibfnamefont {B.}~\bibnamefont
			{Depalle}}, \bibinfo {author} {\bibfnamefont {Z.}~\bibnamefont {Qin}},
		\bibinfo {author} {\bibfnamefont {S.~J.}\ \bibnamefont {Shefelbine}},\ and\
		\bibinfo {author} {\bibfnamefont {M.~J.}\ \bibnamefont {Buehler}},\
	}\bibfield  {title} {\bibinfo {title} {Influence of cross-link structure,
			density and mechanical properties in the mesoscale deformation mechanisms of
			collagen fibrils},\ }\href@noop {} {\bibfield  {journal} {\bibinfo  {journal}
			{Journal of the mechanical behavior of biomedical materials}\ }\textbf
		{\bibinfo {volume} {52}},\ \bibinfo {pages} {1} (\bibinfo {year}
		{2015})}\BibitemShut {NoStop}%
	\bibitem [{\citenamefont {Malaspina}\ \emph {et~al.}(2017)\citenamefont
		{Malaspina}, \citenamefont {Szleifer},\ and\ \citenamefont
		{Dhaher}}]{MALASPINA2017549}%
	\BibitemOpen
	\bibfield  {author} {\bibinfo {author} {\bibfnamefont {D.~C.}\ \bibnamefont
			{Malaspina}}, \bibinfo {author} {\bibfnamefont {I.}~\bibnamefont
			{Szleifer}},\ and\ \bibinfo {author} {\bibfnamefont {Y.}~\bibnamefont
			{Dhaher}},\ }\bibfield  {title} {\bibinfo {title} {Mechanical properties of a
			collagen fibril under simulated degradation},\ }\href
	{https://doi.org/https://doi.org/10.1016/j.jmbbm.2017.08.020} {\bibfield
		{journal} {\bibinfo  {journal} {Journal of the Mechanical Behavior of
				Biomedical Materials}\ }\textbf {\bibinfo {volume} {75}},\ \bibinfo {pages}
		{549} (\bibinfo {year} {2017})}\BibitemShut {NoStop}%
	\bibitem [{\citenamefont {Shen}\ \emph {et~al.}(2008)\citenamefont {Shen},
		\citenamefont {Dodge}, \citenamefont {Kahn}, \citenamefont {Ballarini},\ and\
		\citenamefont {Eppell}}]{SHEN2008cyclicload}%
	\BibitemOpen
	\bibfield  {author} {\bibinfo {author} {\bibfnamefont {Z.~L.}\ \bibnamefont
			{Shen}}, \bibinfo {author} {\bibfnamefont {M.~R.}\ \bibnamefont {Dodge}},
		\bibinfo {author} {\bibfnamefont {H.}~\bibnamefont {Kahn}}, \bibinfo {author}
		{\bibfnamefont {R.}~\bibnamefont {Ballarini}},\ and\ \bibinfo {author}
		{\bibfnamefont {S.~J.}\ \bibnamefont {Eppell}},\ }\bibfield  {title}
	{\bibinfo {title} {Stress-strain experiments on individual collagen
			fibrils},\ }\href
	{https://doi.org/https://doi.org/10.1529/biophysj.107.124602} {\bibfield
		{journal} {\bibinfo  {journal} {Biophysical Journal}\ }\textbf {\bibinfo
			{volume} {95}},\ \bibinfo {pages} {3956} (\bibinfo {year}
		{2008})}\BibitemShut {NoStop}%
	\bibitem [{\citenamefont {Svensson}\ \emph {et~al.}(2010)\citenamefont
		{Svensson}, \citenamefont {Hassenkam}, \citenamefont {Hansen},\ and\
		\citenamefont {{Peter Magnusson}}}]{SVENSSON2010}%
	\BibitemOpen
	\bibfield  {author} {\bibinfo {author} {\bibfnamefont {R.~B.}\ \bibnamefont
			{Svensson}}, \bibinfo {author} {\bibfnamefont {T.}~\bibnamefont {Hassenkam}},
		\bibinfo {author} {\bibfnamefont {P.}~\bibnamefont {Hansen}},\ and\ \bibinfo
		{author} {\bibfnamefont {S.}~\bibnamefont {{Peter Magnusson}}},\ }\bibfield
	{title} {\bibinfo {title} {Viscoelastic behavior of discrete human collagen
			fibrils},\ }\href
	{https://doi.org/https://doi.org/10.1016/j.jmbbm.2009.01.005} {\bibfield
		{journal} {\bibinfo  {journal} {Journal of the Mechanical Behavior of
				Biomedical Materials}\ }\textbf {\bibinfo {volume} {3}},\ \bibinfo {pages}
		{112} (\bibinfo {year} {2010})}\BibitemShut {NoStop}%
	\bibitem [{\citenamefont {Liu}\ \emph {et~al.}(2018)\citenamefont {Liu},
		\citenamefont {Das}, \citenamefont {Yang}, \citenamefont {Schwartz},
		\citenamefont {Genin}, \citenamefont {Thomopoulos},\ and\ \citenamefont
		{Chasiotis}}]{LIU2018cyclicload}%
	\BibitemOpen
	\bibfield  {author} {\bibinfo {author} {\bibfnamefont {J.}~\bibnamefont
			{Liu}}, \bibinfo {author} {\bibfnamefont {D.}~\bibnamefont {Das}}, \bibinfo
		{author} {\bibfnamefont {F.}~\bibnamefont {Yang}}, \bibinfo {author}
		{\bibfnamefont {A.~G.}\ \bibnamefont {Schwartz}}, \bibinfo {author}
		{\bibfnamefont {G.~M.}\ \bibnamefont {Genin}}, \bibinfo {author}
		{\bibfnamefont {S.}~\bibnamefont {Thomopoulos}},\ and\ \bibinfo {author}
		{\bibfnamefont {I.}~\bibnamefont {Chasiotis}},\ }\bibfield  {title} {\bibinfo
		{title} {Energy dissipation in mammalian collagen fibrils: Cyclic
			strain-induced damping, toughening, and strengthening},\ }\href
	{https://doi.org/https://doi.org/10.1016/j.actbio.2018.09.027} {\bibfield
		{journal} {\bibinfo  {journal} {Acta Biomaterialia}\ }\textbf {\bibinfo
			{volume} {80}},\ \bibinfo {pages} {217} (\bibinfo {year} {2018})}\BibitemShut
	{NoStop}%
	\bibitem [{\citenamefont {Elbanna}\ and\ \citenamefont
		{Carlson}(2013)}]{elbanna2013dynamics}%
	\BibitemOpen
	\bibfield  {author} {\bibinfo {author} {\bibfnamefont {A.~E.}\ \bibnamefont
			{Elbanna}}\ and\ \bibinfo {author} {\bibfnamefont {J.~M.}\ \bibnamefont
			{Carlson}},\ }\bibfield  {title} {\bibinfo {title} {Dynamics of polymer
			molecules with sacrificial bond and hidden length systems: towards a
			physically-based mesoscopic constitutive law},\ }\href@noop {} {\bibfield
		{journal} {\bibinfo  {journal} {PloS one}\ }\textbf {\bibinfo {volume} {8}},\
		\bibinfo {pages} {e56118} (\bibinfo {year} {2013})}\BibitemShut {NoStop}%
	\bibitem [{\citenamefont {Lieou}\ \emph {et~al.}(2013)\citenamefont {Lieou},
		\citenamefont {Elbanna},\ and\ \citenamefont {Carlson}}]{Lieou2013pre}%
	\BibitemOpen
	\bibfield  {author} {\bibinfo {author} {\bibfnamefont {C.~K.~C.}\
			\bibnamefont {Lieou}}, \bibinfo {author} {\bibfnamefont {A.~E.}\ \bibnamefont
			{Elbanna}},\ and\ \bibinfo {author} {\bibfnamefont {J.~M.}\ \bibnamefont
			{Carlson}},\ }\bibfield  {title} {\bibinfo {title} {Sacrificial bonds and
			hidden length in biomaterials: A kinetic constitutive description of strength
			and toughness in bone},\ }\href {https://doi.org/10.1103/PhysRevE.88.012703}
	{\bibfield  {journal} {\bibinfo  {journal} {Phys. Rev. E}\ }\textbf {\bibinfo
			{volume} {88}},\ \bibinfo {pages} {012703} (\bibinfo {year}
		{2013})}\BibitemShut {NoStop}%
	\bibitem [{\citenamefont {Rief}\ \emph {et~al.}(1998)\citenamefont {Rief},
		\citenamefont {Fernandez},\ and\ \citenamefont {Gaub}}]{Reif1998}%
	\BibitemOpen
	\bibfield  {author} {\bibinfo {author} {\bibfnamefont {M.}~\bibnamefont
			{Rief}}, \bibinfo {author} {\bibfnamefont {J.~M.}\ \bibnamefont
			{Fernandez}},\ and\ \bibinfo {author} {\bibfnamefont {H.~E.}\ \bibnamefont
			{Gaub}},\ }\bibfield  {title} {\bibinfo {title} {Elastically coupled
			two-level systems as a model for biopolymer extensibility},\ }\href
	{https://doi.org/10.1103/PhysRevLett.81.4764} {\bibfield  {journal} {\bibinfo
			{journal} {Phys. Rev. Lett.}\ }\textbf {\bibinfo {volume} {81}},\ \bibinfo
		{pages} {4764} (\bibinfo {year} {1998})}\BibitemShut {NoStop}%
	\bibitem [{\citenamefont {Su}\ and\ \citenamefont {Purohit}(2009)}]{SU2009}%
	\BibitemOpen
	\bibfield  {author} {\bibinfo {author} {\bibfnamefont {T.}~\bibnamefont
			{Su}}\ and\ \bibinfo {author} {\bibfnamefont {P.~K.}\ \bibnamefont
			{Purohit}},\ }\bibfield  {title} {\bibinfo {title} {Mechanics of forced
			unfolding of proteins},\ }\href
	{https://doi.org/https://doi.org/10.1016/j.actbio.2009.01.038} {\bibfield
		{journal} {\bibinfo  {journal} {Acta Biomaterialia}\ }\textbf {\bibinfo
			{volume} {5}},\ \bibinfo {pages} {1855} (\bibinfo {year} {2009})}\BibitemShut
	{NoStop}%
	\bibitem [{\citenamefont {Bell}(1978)}]{Bell}%
	\BibitemOpen
	\bibfield  {author} {\bibinfo {author} {\bibfnamefont {G.~I.}\ \bibnamefont
			{Bell}},\ }\bibfield  {title} {\bibinfo {title} {Models for the specific
			adhesion of cells to cells: a theoretical framework for adhesion mediated by
			reversible bonds between cell surface molecules.},\ }\href@noop {} {\bibfield
		{journal} {\bibinfo  {journal} {Science}\ }\textbf {\bibinfo {volume}
			{200}},\ \bibinfo {pages} {618} (\bibinfo {year} {1978})}\BibitemShut
	{NoStop}%
	\bibitem [{\citenamefont {Veres}\ \emph {et~al.}(2013)\citenamefont {Veres},
		\citenamefont {Harrison},\ and\ \citenamefont {Lee}}]{veres2013cross}%
	\BibitemOpen
	\bibfield  {author} {\bibinfo {author} {\bibfnamefont {S.~P.}\ \bibnamefont
			{Veres}}, \bibinfo {author} {\bibfnamefont {J.~M.}\ \bibnamefont
			{Harrison}},\ and\ \bibinfo {author} {\bibfnamefont {J.~M.}\ \bibnamefont
			{Lee}},\ }\bibfield  {title} {\bibinfo {title} {Cross-link stabilization does
			not affect the response of collagen molecules, fibrils, or tendons to tensile
			overload},\ }\href@noop {} {\bibfield  {journal} {\bibinfo  {journal}
			{Journal of Orthopaedic Research}\ }\textbf {\bibinfo {volume} {31}},\
		\bibinfo {pages} {1907} (\bibinfo {year} {2013})}\BibitemShut {NoStop}%
	\bibitem [{\citenamefont {Linka}\ and\ \citenamefont
		{Itskov}(2016)}]{linka2016mechanics}%
	\BibitemOpen
	\bibfield  {author} {\bibinfo {author} {\bibfnamefont {K.}~\bibnamefont
			{Linka}}\ and\ \bibinfo {author} {\bibfnamefont {M.}~\bibnamefont {Itskov}},\
	}\bibfield  {title} {\bibinfo {title} {Mechanics of collagen fibrils: A
			two-scale discrete damage model},\ }\href@noop {} {\bibfield  {journal}
		{\bibinfo  {journal} {Journal of the mechanical behavior of biomedical
				materials}\ }\textbf {\bibinfo {volume} {58}},\ \bibinfo {pages} {163}
		(\bibinfo {year} {2016})}\BibitemShut {NoStop}%
	\bibitem [{\citenamefont {Plimpton}(1995)}]{LAMMPS}%
	\BibitemOpen
	\bibfield  {author} {\bibinfo {author} {\bibfnamefont {S.}~\bibnamefont
			{Plimpton}},\ }\bibfield  {title} {\bibinfo {title} {Fast parallel algorithms
			for short-range molecular dynamics},\ }\href@noop {} {\bibfield  {journal}
		{\bibinfo  {journal} {Journal of computational physics}\ }\textbf {\bibinfo
			{volume} {117}},\ \bibinfo {pages} {1} (\bibinfo {year} {1995})}\BibitemShut
	{NoStop}%
\end{thebibliography}
%

\end{document}